\documentclass[a4paper,12pt]{article}
\usepackage{amsmath,amsfonts}
\def\ba {\begin {array}}
\def\ea{\end {array}}
\def\be {\begin {equation}}
\def\ee {\end {equation}}
\def\bea {\begin {eqnarray}}
\def\eea {\end {eqnarray}}

\begin{document}
\title
{Structure results for higher order symmetry algebras of 2D classical superintegrable systems}

\author{E. G.~Kalnins\\
Department of Mathematics,\\
University
of Waikato, Hamilton, New Zealand.\\
and\\
W.~Miller, Jr.
 School of Mathematics, University of Minnesota,\\
Minneapolis, Minnesota, U.S.A.}
\maketitle

MSC classes: 20C99, 20C35, 22E70
\begin{abstract} Recently the  authors and J.M. Kress presented a special function recurrence relation method to prove quantum
superintegrability of an integrable  2D system that included  explicit
constructions of  higher order  symmetries and the structure relations for the closed algebra generated by these symmetries. 
We applied the method to 5 families of systems, each depending on a rational parameter $k$, including
most notably the caged anisotropic oscillator, the  Tremblay, 
Turbiner and Winternitz system and a deformed Kepler-Coulomb system. Here we work out the analogs of these constructions for all of 
the associated classical Hamiltonian systems, as well as for a family including the generic potential on the 2-sphere. We do not have
 a proof in every case that the generating symmetries are of lowest possible order, but we believe this to be so via an extension of our
 method.  
\end{abstract}


\section{ Introduction}\label{int1}
Recently Tremblay, Turbiner and Winternitz \cite{TTW1, TTW2}  studied a  family
of quantum and classical mechanical systems in two dimensions with classical
Hamiltonian 
\be\label{KMP1}
{\cal H}=p^2_r+\frac{1}{ r^2}p^2_\theta -\omega^2r^2+ \frac{\alpha}{ r^2\cos ^2k\theta } 
+\frac{\beta}{r^2\sin ^2k\theta }.\ee
We call this the TTW system. They conjectured  and gave strong evidence that for  $k$  a rational number 
 this system is superintegrable, as is the corresponding quantum analog

\be\label{TTWham}
H=\partial ^2_r+\frac{2}{ r}\partial _r+\frac{1}{ r^2}\partial ^2_\theta -\omega^2r^2+ 
\frac{\alpha}{ r^2\cos ^2k\theta } +\frac{\beta}{ r^2\sin ^2k\theta }.\ee
 In papers \cite{KKM10,KMPog10} the authors and collaborators proved
quantum and classical superintegrability for these systems, the first proofs that covered  all rational $k$. Related results are
 \cite{KKM10a,KKM10b,KKM10c}.
This system has sparked a great deal of interest because it includes several special cases of definite physical interest, 
it provides many examples
of superintegrability where the generating constants of the motion are of very high order as polynomials in the momenta or 
as differential operators in the quantum case, and because it 
gave a hint as to how many other such families of classical and quantum systems could be constructed. The previous classical proofs of 
superintegrability for these systems have usually not determined the structure of the symmetry algebras, or even 
verified closure at finite order under commutation. For 2D quantum systems, however, the authors and J. M. Kress have recently 
introduced a method for verifying superintegrability based on recurrence relations for hypergeometric functions that enables one to 
compute the associated  structure relations for the symmetry algebra with relative ease. In \cite{KKM10c} structure relations were obtained for 5 
families of superintegrable systems, each indexed by a rational parameter $k$, and including the TTW family. 
Here we give a simplified exposition of some of these results. 

Each  of these quantum systems  in 2D has an analogous classical counterpart that can be shown to be  classically superintegrable. Natural questions are 
\begin{itemize}\item What is the structure of the classical symmetry algebra for each of these systems? Does it close at finite order?
 \item How are the structures of the classical and quantum symmetry algebras, and the symmetries themselves, related.
\end{itemize}
We will present new results in this paper that provide at least partial answers  to these questions for each of the systems
 studied in \cite{KKM10c}, as well as for a family that contains the 2D system on the 2-sphere with generic potential.

\section{Brief resume of definitions}
Suppose we have a Hamiltonian system on a 2D local Riemannian manifold (real or complex) such that the Hamilton-Jacobi equation admits 
additive separation in some orthogonal 
coordinate system. Then the separable coordinates $x_1,x_2$ can always be chosen such that the Hamiltonian takes the form
\be\label{sepconst2} {\cal H}={\cal L}_1=\frac{1}{f_1(x_1)+f_2(x_2)}(p_{x_1}^2+p_{x_2}^2+v_1(x_1)+v_2(x_2),\ee
and the phase space function 
$$  {\cal  L}_2 =\frac{f_2(x_2)}{f_1(x_1)+f_2(x_2)}\left(p_{x_1}^2+v_1(x_1)\right)-\frac{f_1(x_1)}{f_1(x_1)+f_2(x_2)}\left(p_{x_2}^2+v_2(x_2)\right).$$ 
is the second order constant of the motion determining the separation. Here, $\{{\cal  L}_2,{\cal H}\}=0$ where 
 $$\{{\cal F}, {\cal G}\}=\sum_{j=1}^2\left(\partial_{x_j}{\cal F}\partial_{p_{x_j}}{\cal G}-\partial_{p_{x_j}}{\cal F}\partial_{x_j}{\cal G}\right)$$
is the Poisson bracket of two phase space functions ${\cal F},{\cal G}$. A constant of the motion is a nonzero function on the phase space 
whose Poisson bracket with
 $\cal H$ vanishes. Clearly $\cal H$ and ${\cal L}_2$ are functionally independent constants of the motion, second order in the momenta. 
We will present strategies for determining functions $f_1,f_2,v_1,v_1$ such that there exists a 3rd constant of the motion, polynomial in the momenta,
 and such that 
the three constants are functionally independent. In this case the system is superintegrable.
By what is essentially an action angle variable construction \cite{KKM10b}, one can compute the integrals $M(x)=\frac12\int\frac{dx}{p_x}$ and 
$N(y)=\frac12\int\frac{dy}{p_y}$. Then  ${\tilde {\cal L}}_2=M-N$ is a third constant of the motion, but usually not a polynomial in the momenta, 
hence not directly useful in verifying superintegrability. In \cite{KMPog10,KKM10b}, see also \cite{CG, MPT}, we described a procedure for obtaining a 
polynomial constant from $M-N$, based on the observation that the integrals
 $$M=\frac12\int \frac{dx_1}{\sqrt{f_1{\cal H}+{\cal L}_2-v_1}},\quad N=\frac12\int \frac{dx_2}{\sqrt{f_2{\cal H}-{\cal L}_2-v_2}},$$
can often be expressed in terms of multiples of the inverse hyperbolic sine or cosine (or the ordinary inverse sine or cosine), and the 
hyperbolic sine and cosine satisfy addition formulas. Our method works for  functions $f_j,v_j$ such that $M$ and $N$ possess this property. 

For 2D quantum systems such that the Schr\"odinger eigenvalue equation $H\Psi=E\Psi$ admits multiplicative separation in an 
orthogonal coordinate system we have 
\be\label{sepconst2Q} { H}={ L}_1=\frac{1}{f_1(x_1)+f_2(x_2)}(\partial_{x_1}^2+\partial_{x_2}^2+v_1(x_1)+v_2(x_2),\ee
and 
$$  {  L}_2 =\frac{f_2(x_2)}{f_1(x_1)+f_2(x_2)}\left(\partial_{x_1}^2+v_1(x_1)\right)-\frac{f_1(x_1)}{f_1(x_1)+f_2(x_2)}\left(\partial_{x_2}^2+v_2(x_2)\right).$$ 
is the second order operator  determining the separation. Here, $[  L_2, H]=0$ where 
 $[ F, G]=FG-GF$ is the commutator 
 of two operators $ F,G$. The separable eigenfunctions $\Psi=\Phi(x_1)\Theta(x_2)$ of $H$ are then characterized by the equations
$$ H\Psi=E\Psi,\quad L_2\Psi=\lambda \Psi$$
where $E$ is the energy eigenvalue and $\lambda$ is the separation constant. A symmetry operator is a nonzero differential operator in $x_1,x_2$
whose commutator with
 $ H$ vanishes. Clearly the second order differential operators  $ H$ and $ L_2$ are algebraically independent symmetries. In \cite{KKM10c}
we presented a recurrence relation method  for determining  a 3rd finite order symmetry operator  such that the three symmetries are algebraically 
independent, i.e. such that the system is quantum superintegrable.

\section{The quantum TTW system} We use the quantum TTW system to illustrate our recurrence relation approach to proving superintegrability and determining the structure
of the symmetry algebra. For complete details on this construction and on its application to other systems, see  \cite{KKM10c}.
 The quantum Hamiltonian operator is 
(\ref{TTWham}).The general solution of the eigenvalue 
problem $H\Psi =E\Psi $ is 
\be\label{sepeigenfunctions}\Psi  = e^{-\frac{\omega }{ 2}r^2}r^{k(2n+a+b+1)} 
L^{k(2n+a+b+1)}_m(\omega r^2)
(\sin (k\theta ))^{a+\frac{1}{ 2}}(\cos (k\theta ))^{b+\frac{1}{ 2}} 
P^{a,b}_n(\cos (2k\theta ))\ee
where we have taken $\alpha =k^2(\frac{1}{ 4} -a^2)$ and $\beta =k^2(\frac{1}{
4} 
-b^2)$. The $L$-functions are associated Laguerre and the $P$-functions are
Jacobi, 
not polynomials in general, \cite{AAR}.   We consider the functions 
$$\Pi  = e^{-\frac{\omega }{2}r^2}r^{k(2n+a+b+1)} 
L^{k(2n+a+b+1)}_m(\omega r^2)P^{a,b}_n(x)=Y^{ A}_m(r)X^{a,b}_n(x)$$
where $x=\cos (2k\theta )$,  $X=P$ or $Q$ and $Y=S$ or $T$. (Here $\Pi$ is obtained from $\Psi$ by a
gauge transformation to remove the angular factors
$(\sin (k\theta ))^{a+\frac{1}{ 2}}(\cos (k\theta ))^{b+\frac{1}{ 2}}$.)
 By this we mean that $P^{a,b}_n(x)$ is a 
Jacobi polynomial if $n$ is an integer. Otherwise it is given by its 
hypergeometric expression. If $X=Q$ then this denotes the associated second 
solution of the Jacobi differential equation. Similar remarks apply to the 
choice of $Y=S$.  We  have defined this function to be  
$$S^{ A}_m(r)=e^{-\frac{\omega }{ 2}r^2}r^{k(2n+a+b+1)} 
L^{k(2n+a+b+1)}_m(\omega r^2)$$
and $T^{ A}_m(r)$ to be a second independent solution. The energy eigenvalue
is given 
by  
\be\label{energyev}E=-2\omega \left[2(m+nk)+1+(a+b+1)k\right]\ee
and $ A=k(2n+a+b+1)$.
The separation equation for $\Theta (\theta )$ is 
$${\tilde L}_2\Theta=(\partial ^2_\theta + \frac{\alpha }{ \sin ^2(k\theta )} + 
\frac{\beta }{\cos ^2(k\theta )})\Theta (\theta )=-k^2(2n+a+b+1)^2\Theta (
\theta )=-A^2 \Theta (\theta )$$
and ${\tilde L}_2$ is a symmetry operator for  the system. Under the gauge
transformation ${\tilde L}_2$  goes to a symmetry 
that we shall  call $L_2$ and which has the same eigenvalues.
We see from the expression for $E$ that in order that an energy eigenvalue to be
unchanged for different values of $m,n$ 
 we must fix $u=m+nk$. The   two transformations  
$$
n\rightarrow n+q,\quad m\rightarrow m-p$$
and  
$$n\rightarrow n-q,\quad m\rightarrow m+p$$
will each achieve this.

Consider the functions $X^{a,b}_n(x)$. If we want to raise or lower
the index $n$ we can do so with the operators 
$$J^+_n X^{a,b}_n(x)=$$
$$(2n+a+b+2)(1-x^2)\partial
_xX^{a,b}_n(x)+(n+a+b+1)(-(2n+a+b+2)x-(a-b))X^{a,b}_n(x)$$
$$=2(n+1)(n+a+b+1)X^{a,b}_{n+1}(x)$$
and 
$$J^-_n X^{a,b}_n(x)=$$
$$-(2n+a+b)(1-x^2)\partial
_xX^{a,b}_n(x)-n((2n+a+b)x-(a-b))X^{a,b}_n(x)=2(n+a)(n+b)X^{a,b}_{n-1}(x),$$
\cite{AAR}.
Similarly, for the  functions ${\cal Y}^A_m(R)=\omega^{A/2}Y^{A}_m(r)$ where
$R= r^2$ we can deduce the 
relations 
$$K^+_{A,m}{\cal Y}^A_m(R)=\left\{(A+1)\partial _R-\frac{E}{ 4 }-\frac{1}{
2R}A(A+1)\right\}{\cal Y}^A_m(R)
= -\omega{\cal Y}^{A+2}_{m-1}(R),$$
$$K^-_{A,m}{\cal Y}^A_m(R)=\left\{(-A+1) 
\partial _R-\frac{E}{ 4}+\frac{1}{ 2R}A(1-A)\right\}{\cal Y}^A_m(R)
=-\omega(m+1)(m+A) {\cal Y}^{A-2}_{m+1}(R),$$ \cite{AAR}.
 We now  construct the two operators
\be\label{recurrence3} \Xi _+=K^+_{A+2(p-1),m-(p-1)}\cdots
K^+_{A,m}J^+_{n+q-1}\cdots J^+_n\ee
and  
\be\label{recurrence4}\Xi _-=K^-_{A-2(p-1),m+p-1}\cdots
K^-_{A,m}J^-_{n-q+1}\cdots J^-_n. \ee
When applied to a basis function $\Psi_n={\cal Y}^A_m(R)X^{a,b}_n(x)$ for fixed
$u=m+kn$, (so $m=u-kn$)
these operators  raise and lower indices according to
\be\label{raise1} \Xi _+\Psi_n=2^q(-1)^p \omega^p(n+1)_q(n+a+b+1)_q\Psi_{n+q},\ee
\be\label{lower1} \Xi _-\Psi_n=2^q \omega^p
(-n-a)_q(-n-b)_q(u-kn+1)_p(-u-k[n+a+b+1])_p\Psi_{n-q},\ee
where $(\alpha)_q=(\alpha)(\alpha+1)\cdots(\alpha+q-1)$ for nonnegative integer
$q$. It is easy to verify that 
under the transformation $n \rightarrow  -n-a-b-1$  we have ${\Xi_+} \rightarrow
\Xi_-$ and ${\Xi_-} \rightarrow \Xi_+$. 
Thus $\Xi=\Xi_++\Xi_-$  as a polynomial in $n$ and $u$ is unchanged under this
transformation. 
Therefore it is a polynomial in $(2n+a+b+1)^2$ and $u$.  As a consequence of 
the relation $\lambda =-k^2(2n+a+b+1)^2$,   in the
expansion of 
$\Xi$ in terms of powers of $(2n+a+b+1)^2$ and $E$ we can replace $(2n+a+b+1)^2$
by $L_2/k^2$ and $E$ by $H$ everywhere they occur, and express $\Xi$ as a
pure differential operator, independent of parameters.  Thus, in the
expansion of $\Xi$ in terms of the parameters, a term $ W E$
 is replaced by 
$WH$
with the $W$ operator on the left.
Clearly this operator, which we will also call $\Xi$ commutes with $H$ on the formal
eigenspaces of $H$. However, in  \cite{KKM10c} we showed  shows that in fact $\Xi$ commutes with $H$ in general,
thus it is a symmetry operator for $H$.(Indeed it is shown there that any polynomial operator
 identity established on all formal eigenspaces must hold identically.This is a general fact that holds for all 
applications of the recurrence operator method.)

We can also easily see that under the transformation $n \rightarrow  -n-a-b-1$  
the operator $\Xi_+ -\Xi_-$ changes sign, hence the operator 
 ${\tilde \Xi}=(1/(2n+a+b+1))(\Xi_+ -\Xi_-)$ is unchanged under this
transformation. Again, making the replacements $(2n+a+b+1)^2$ 
by $L_2/k^2$ and $u$ by 
$-2 -(H+2\omega k(a+b+1))/4\omega$ we can express ${\tilde \Xi}$ as a pure
differential operator, independent of parameters, 
and it is a symmetry operator for $H$. Each of these symmetries has a nonzero
commutator with $L_2$, so each is 
algebraically independent of the set $H,L_2$. This proves that the TTW system is
quantum superintegrable for all rational $k$.
We set $L_3= \Xi$, $L_4={\tilde \Xi}$.

Using the explicit relations (\ref{raise1}), (\ref{lower1}) for the action of
the raising and lowering operators $\Xi_\pm$ on a basis
we can obtained very detailed information about the structure of the symmetry
algebra generated by $L_2,L_3,L_4$. 
Applying the raising operator to a basis function, followed by the lowering
operator, we obtain the result
\be\label{LR1}  \Xi_-\Xi_+\Psi_n\ee
$$=(-1)^p4^q\omega^{2p}(n+1)_q(n+a+1)_q(n+b+1)_q(n+a+b+1)_q(-u+kn)_p(u+k[n+a+b+1]
+1)_p\Psi_n$$
$$=\xi_n\Psi_n,$$
Reversing the order we find
\be\label{RL1} \Xi_+\Xi_-\Psi_n\ee
$$=(-1)^p4^q\omega^{2p}(-n)_q(-n-a)_q(-n-b)_q(-n-a-b)_q(u-kn+1)_p(-u-k[n+a+b+1])_p\Psi_n$$
$$ =\eta_n\Psi_n.$$
Thus the action of the operator $\Xi_-\Xi_++ \Xi_+)\Xi_-$ on any
basis function $\Psi_n$ is to multiply it by 
$\xi_n+\eta_n$. However, it is easy to check from expressions (\ref{raise1}),
(\ref{lower1} and from  (\ref{RL1}), (\ref{LR1}) 
that under the transformation $n \rightarrow  -n-a-b-1$  we have  
$\Xi_-\Xi_+\leftrightarrow \Xi_+\Xi_-$ and   
  $\xi_n\leftrightarrow \eta_n$. Thus $\Xi^{(+)}=\Xi_-\Xi_++
\Xi_+\Xi_-$ is an even polynomial operator in $(2n+a+b+1)$, 
polynomial in $u$, and 
 $\xi_n+\eta_n$ is an even polynomial function in $(2n+a+b+1)$, polynomial in
$u$. Furthermore, each of $\Xi_-\Xi_+$ and $\Xi_+\Xi_-$ is unchanged under the transformation
$u\longrightarrow -u-(a+b+1)-1$, hence each is a polynomial of order $p$ in 
$[2u+(a+b+1)k+1]^2=E^2/4\omega^2$. Due to the multiplicative factor $\omega^{2p}$ in each of these expressions
we conclude that $\Xi^{(+)}$ is a 
symmetry operator whose action on a basis is
 given by a polynomial operator $P^{(+)}(H^2,L_2,\omega^2,a,b)$. In fact, 
$$ \Xi^{(+)}=P^{(+)}(H^2,L_2,\omega^2,a,b).$$

Similarly, the operator $\Xi^{(-)}=(\Xi_-\Xi_+-
\Xi_+\Xi_-)/(2n+a+b+1)$ is an even polynomial 
in $(2n+a+b+1)$, as is $(\xi_n-\eta_n)/(2n+a+b+1)$. Also it is polynomial in $E^2$ and $\omega^2$. Thus
$\Xi^{(-)}=P^{(-)}(H^2,L_2,\omega^2,a,b)$ is a symmetry operator 
which is a polynomial function of all of its variables.

Now we can compute the structure relations explicitly by evaluating the
operators on an eigenfunction basis. 
These relations must then hold everywhere. The results are:
\be\label{[L2,L4]} [L_2,L_4]=-4k^2qL_3-4k^2q^2L_4,\ee
\be\label{[L2,L3]} [L_2,L_3]=2q\{L_2,L_4\}+4k^2q^2L_3+8k^2q^3L_4,\ee
\be\label{[L3,L4]} [L_3,L_4]=2qL_4^2-2P^{(-)}(H^2,L_2,\omega^2,a,b),\ee
\be\label{L3^2Q}
6k^2L_3^2+\{L_2,L_4,L_4\}+6k^2q\{L_3,L_4\}+28k^2q^2L_4^2-4k^2qP^{(-)}(H^2,
L_2,\omega^2,a,b)\ee
$$-12k^2P^{(+)}(H^2,L_2,\omega^2,a,b)=0.$$
Here, $\{A,B\}=AB+BA$ and $\{A,B,C\}$ is the analogous  6-term symmetrizer of 3
operators.
A more transparent realization, for $R=-4k^2qL_3-4k^2q^2L_4$, is
\be\label{[L2,L4R]} [L_2,L_4]=R,\ee
\be\label{[L2,R]} [L_2,R]=-8k^2q^2\{L_2,L_4\}-16k^4q^4L_4,\ee
\be\label{[L4,R]} [L_4,R]=8k^2q^2L_4^2-8k^2qP^{(-)}(H^2,L_2,\omega^2,a,b),\ee
\be\label{R^2}
\frac{3}{8k^2q^2}R^2+22k^2q^2L_4^2+\{L_2,L_4,L_4\}-4k^2qP^{(-)}(H^2,L_2,\omega^2,a,b)\ee
$$-12k^2P^{
(+)}(H^2,L_2,\omega^2,a,b)=0.$$
From this result we see that the symmetries $H,L_2,L_4$ generate a closed
symmetry algebra. 

For $k=1$ the operator is of order 3, whereas we know that there is a 2nd order symmetry, due to the fact that the TTW Hamiltonian-Jacobi equation  for $k=1$ is multiply separable. Thus we have produced only a proper subalgebra of the maximal symmetry algebra. However, for $k\ne 1$ this multiple separation no longer occurs and we think that the generators we have constructed are of minimum order.

\section{The classical TTW system}
The TTW Hamiltonian  (\ref{KMP1}) admits a second order constant of the motion 
corresponding to separation of variables in polar coordinates, viz 
\be\label{angmom}
{\cal L}_2=p^2_\theta +\frac{\alpha}{ \cos ^2k\theta } +\frac{\beta} {\sin ^2k\theta }.\ee
To demonstrate superintegrability we need to exhibit a third polynomial constant of the motion 
that is functionally independent of ${\cal H}, {\cal L}_2$. The method for constructing this symmetry is based on 
action angle variables and the existence of an additively separable coordinate system for the Hamilton-Jacobi equation,
 \cite{KMPog10, Evans2008a}. We briefly review the construction.
In terms of the new variable $r=e^R$ the Hamiltonian assumes the form 
$${\cal H}=e^{-2R}(p^2_R+p^2_\theta -\omega^2 e^{4R}+\frac{\alpha }{ \cos ^2(k\theta )}+ 
\frac{\beta }{ \sin ^2(k\theta )}).$$
Applying the method of \cite{KMPog10}
to find  extra invariants we  first need to construct a 
function $M(R,p_R)$ which satisfies $\{M,{\cal H}\}=e^{-2R}$, or
$$(4\omega^2 e^{4R}+2{\cal H}e^{2R})\partial _{p_R}M+2p_R\partial_R M=1.$$
This equation has a solution  
$$M=\frac{i}{ 4\sqrt{ {\cal L}_2}} B$$
where 
$$
\sinh{ B} =i \frac{(2{\cal L}_2e^{-2R}-{\cal H})}{ \sqrt{ {\cal  H}^2+4\omega^2 {\cal  L}_2}},\quad
\cosh{B }= \frac{2\sqrt{ {\cal L}_2}e^{-2R}p_R}{ \sqrt{ {\cal  H}^2+4\omega^2 {\cal L}_2}},$$
and 
\be\label{L2}{\cal L}_2=p^2_\theta +\frac{\alpha }{ \cos ^2(k\theta )}+ 
\frac{\beta }{ \sin ^2(k\theta )},\ee
and we also have the relation (which we can use to consider $M$ as a function of $R$ alone):
\be\label{LHident}p^2_R+{\cal L}_2-\omega^2 e^{4R}-e^{2R}{\cal H}=0.\ee
We now compute a corresponding function $N(\theta ,p_\theta )$ which
 satisfies $\{N,{\cal H}\}=e^{-2R}$, or $$-(\frac{\alpha }{ \cos ^2(k\theta )}+ 
\frac{\beta }{ \sin ^2(k\theta )})'\partial _{p_\theta }N+2p_\theta \partial _\theta N=1$$
where the prime denotes differentiation with respect to $\theta$. This equation has a
solution 
$N= - \frac{i}{ 4\sqrt{{\cal  L}_2}k} A$
where  
$$
\sinh{A }= 
i\frac{-\beta +\alpha -{\cal L}_2\cos (2k\theta )}{ \sqrt{({\cal L}_2-\alpha -\beta )^2-4\alpha\beta }},\quad
\cosh{ A }= 
\frac{\sqrt{ L_2}\sin (2k\theta )p_\theta }{\sqrt{ ({\cal L}_2-\alpha -\beta )^2-4\alpha \beta }}.$$
According to the action angle variable theory  $M-N$ is a constant of the motion and, since it is 
constructed such that $\{M-N,{\cal L}_2\}\ne 0$, it is functionally independent of ${\cal L}_2$, \cite{KKMPog02}.

From these expressions for $M$ and $N$  we see that if $k$ is rational,
$k=\frac{p}{ q}$ ( where $p, q$ are relatively prime integers) then  
$$-4ip\sqrt{{\cal  L}_2}[N-M]=qA+pB$$ and any function of $qA+pB$ is a constant of the motion (but not polynomial in the momenta).
Our original proofs of superintegrability for this system were based on the fact that the hyperbolic functions
 $\sinh x$, $\cosh x$ satisfied addition theorems. 
Here, however, we will make use of  an observation in \cite{CG} that simplifies the analysis and work with the exponential function.
We have 
\be\label{expA} e^A=\cosh A +\sinh A= {X}/{U},\quad e^{-A}=\cosh A -\sinh A= {\overline{X}}/{U},\ee
\be\label{expB} e^B=\cosh B +\sinh B= {Y}/{S},\quad e^{-B}=\cosh B -\sinh B= {\overline{Y}}/{S},\ee
where
$$ X= \sqrt{{\cal L}_2}\sin (2k\theta) p_\theta+i(-\beta+\alpha-{\cal L}_2\cos(2k\theta)),\   
{\overline X}= \sqrt{{\cal L}_2}\sin (2k\theta) p_\theta-i(-\beta+\alpha-{\cal L}_2\cos(2k\theta)),$$
$$Y= 2\sqrt{{\cal L}_2} e^{-2R}p_R+i(2{\cal L}_2e^{-2R}-{\cal H}),\ 
{\overline Y}= 2\sqrt{{\cal L}_2} e^{-2R}p_R-i(2{\cal L}_2e^{-2R}-{\cal H}),$$
$$ U=\sqrt{({\cal L}_2-\alpha-\beta)^2-4\alpha\beta},\quad S=\sqrt{{\cal H}^2+4\omega^2{\cal L}_2}.$$

Now note that $e^{qA+pB}$ and $e^{-(qA+pB)}$ are constants of the motion, where
\be\label{expAB} e^{qA+pB}=(e^A)^q (e^B)^p=\frac{X^qY^p}{U^qS^p},\quad e^{-(qA+pB)}=(e^{-A})^q (e^{-B})^p=\frac{(\overline{X})^q
(\overline{Y})^p}{U^qS^p}.\ee Moreover, the identity $e^{qA+pB}e^{-(qA+pB)}=1$ can be expressed as
\be\label{fundident} X^q(\overline{X})^qY^p(\overline{Y})^p=U^{2q}S^{2p}=P({\cal L}_2, {\cal H})\ee
where $P$ is a polynomial in ${\cal L}_2$ and ${\cal H}$.

We make some observations. Let $a,b,c,d$ be nonzero complex numbers and consider the 
binomial expansion 
\be\label{plusbinom}\frac12\left[(\sqrt{{\cal L}_2}a+ib)^q(\sqrt{{\cal L}_2}c+id)^p+(\sqrt{{\cal L}_2}a-ib)^q(\sqrt{{\cal L}_2}c-id)^p\right] 
\ee
$$=\sum_{0\le \ell\le q, 0\le s\le p}\left(\ba{cc} q\\ \ell\ea\right) 
\left(\ba{cc} p\\ s\ea\right)b^\ell d^s a^{q-\ell}c^{p-s}\frac{{\cal L}_2^{(q+p-\ell-s)/2}}{2}[i^{\ell+s}+(-i)^{\ell+s}].$$
Suppose $p+q$ is odd. Then it is easy to see that the sum (\ref{plusbinom}) takes the form
 $\sqrt{{\cal L}_2}T_{\rm odd}({\cal L}_2) $ where 
$T_{\rm odd}$ is a polynomial in ${\cal L}_2$. On the other hand, if $p+q$ is even then the sum (\ref{plusbinom}) takes the 
form $T_{\rm even}({\cal L}_2) $ where 
$T_{\rm even}$ is a polynomial in ${\cal L}_2$. 

Similarly, consider the 
binomial expansion 
\be\label{minusbinom}\frac{1}{2i}\left[(\sqrt{{\cal L}_2}a+ib)^q(\sqrt{{\cal L}_2}c+id)^p-
(\sqrt{{\cal L}_2}a-ib)^q(\sqrt{{\cal L}_2}c-id)^p\right] 
\ee
$$=\sum_{0\le \ell\le q, 0\le s\le p}\left(\ba{cc} q\\ \ell\ea\right) 
\left(\ba{cc} p\\ s\ea\right)b^\ell d^s a^{q-\ell}c^{p-s}\frac{{\cal L}_2^{(q+p-\ell-s)/2}}{2i}[i^{\ell+s}-(-i)^{\ell+s}].$$
Suppose $p+q$ is odd. Then  the sum (\ref{minusbinom}) takes the form
 $V_{\rm odd}({\cal L}_2) $ where 
$V_{\rm odd}$ is a polynomial in ${\cal L}_2$. On the other hand, if $p+q$ is even then the sum (\ref{minusbinom}) takes the 
form $\sqrt{{\cal L}_2}V_{\rm even}({\cal L}_2) $ where 
$V_{\rm even}$ is a polynomial in ${\cal L}_2$. 

\medskip\noindent {\bf Case $p+q$ odd}:  Let 
\be \label{oddL34} {\cal L}_4=\frac{1}{\sqrt{{\cal L}_2}}({\cal L}^+ +{\cal L}^-),\ 
{\cal L}_3=\frac{1}{i}({\cal L}^+ -{\cal L}^-),\ee
where
\be\label{oddL+-} {\cal L}^+=X^qY^p,\quad {\cal L}^-=(\overline{X})^q(\overline{Y})^p.\ee
Then we see from (\ref{plusbinom}), (\ref{minusbinom}) that ${\cal L}_3,{\cal L}_4$ are constants of the motion,
 polynomial in the momenta. Moreover, the identity (\ref{fundident}) becomes
\be\label{fundident1}{\cal L}^+{\cal L}^-=P({\cal L}_2,{\cal H}).\ee
Here ${\cal L}^\pm$ are constants of the motion, but are not polynomial in the momenta. We will see that they are the classical 
versions of quantum
raising and lowering operators for angular momentum.

It is straightforward to verify the Poisson bracket computations 
\be\label{raising}\{{\cal L}_2,{\cal L}^+\}=q\frac{X^{q-1}Y^p}{U^qS^p}\{{\cal L}_2,X\}=-4ip\sqrt{{\cal L}_2}{\cal L}^+,\ee
\be\label{lowering}\{{\cal L}_2,{\cal L}^-\}=q\frac{(\overline{X})^{q-1}(\overline{Y})^p}{U^qS^p}\{{\cal L}_2,\overline{X}\}
=4ip\sqrt{{\cal L}_2}{\cal L}^-.\ee
From these results we can obtain
$$\{{\cal L}_2,{\cal L}_4\}=\frac{1}{\sqrt{{\cal L}_2}}(-4ip\sqrt{{\cal L}_2}{\cal L}^++ 4ip\sqrt{{\cal L}_2}{\cal L}^-)$$
$$=-4ip{\cal L}^++4ip{\cal L}^-=4p{\cal L}_3,$$
$$\{{\cal L}_2,{\cal L}_3\}= \frac{1}{i}(-4ip\sqrt{{\cal L}_2}{\cal L}^+-4ip\sqrt{{\cal L}_2}{\cal L}^-)$$
$$=-4p{\cal L}_2{\cal L}_4.$$

To get the remaining structure equations we need some more identities. First of all, note that
\be\label{L4L4} {\cal L}_4^2=\frac{1}{{\cal L}_2}\left[ ({\cal L}^+)^2+2{\cal L}^+{\cal L}^-+({\cal L}^-)^2\right],\ee
\be\label{L3L3}{\cal L}_3^2=-\left[ ({\cal L}^+)^2-2{\cal L}^+{\cal L}^-+({\cal L}^-)^2\right].\ee
Thus,
$ {\cal L}_3^2+{\cal L}_2{\cal L}_4^2=4{\cal L}^+{\cal L}^-=4P({\cal L}_2,{\cal H})$, so
\be\label{L3^2} {\cal L}_3^2=-{\cal L}_2{\cal L}_4^2+4P({\cal L}_2,{\cal H}).\ee

Further,  
$$\{{\cal L}^+,{\cal L}^-\} =\{{\cal L}^+, \frac{P}{{\cal L}^+}\}=  ({\cal L}^+)^{-1}\{{\cal L}^+,P\}=({\cal L}^+)^{-1}\left(\frac{\partial P}{\partial {\cal L}_2}\{{\cal L}^+,{\cal L}_2\}+\frac{\partial P}{\partial {\cal H}}\{{\cal L}^+,{\cal H}\}\right),$$
so
\be\label{L+L-} \{{\cal L}^+,{\cal L}^-\}=4ip\sqrt{{\cal L}_2}\frac{\partial P}{\partial{\cal L}_2}.\ee
It remains only to evaluate $\{{\cal L}_3,{\cal L}_4\}$. We have
$$\{{\cal L}_3,{\cal L}_4\}=\frac{1}{i}\{{\cal L}^+-{\cal L}^-,\frac{{\cal L}^+}{\sqrt{{\cal L}_2}}+\frac{{\cal L}^-}{\sqrt{{\cal L}_2}}\}=$$
$$\frac{1}{i}\left[-({\cal L}_2)^{-3/2}\{{\cal L}^+,{\cal L}_2\}{\cal L}^+ -({\cal L}_2)^{-3/2}\{{\cal L}^+,{\cal L}_2\}{\cal L}^- 
+({\cal L}_2)^{-3/2}\{{\cal L}^-,{\cal L}_2\}{\cal L}^++({\cal L}_2)^{-3/2}\{{\cal L}^-,{\cal L}_2\}{\cal L}^-\right.$$
$$\left. +({\cal L}_2)^{-1/2}\{{\cal L}^+,{\cal L}^-\}-({\cal L}_2)^{-1/2}\{{\cal L}^-,{\cal L}^+\}\right]$$
$$=-\frac{4p}{{\cal L}_2}[({\cal L}^+)^2+2{\cal L}^+{\cal L}^-+({\cal L}^-)^2]+8p\frac{\partial P}{\partial{\cal L}_2}.$$
Then, using (\ref{L4L4}), we conclude that 
\be\label{L3L4bracket}\{{\cal L}_3,{\cal L}_4\}=-4p{\cal L}_4^2+8p\frac{\partial P}{\partial{\cal L}_2}.\ee
To list the structure equations in standard form we define ${\cal R}=4p{\cal L}_3$. Then the structure equations take the form
\be\label{structure+} \{{\cal L}_2,{\cal L}_4\}={\cal R},\ \{{\cal L}_2,{\cal R}\}=-16p^2{\cal L}_2{\cal L}_4,\ \{{\cal L}_4,{\cal R}\}=16p^2{\cal L}_4^2-32p^2\frac{\partial P}{\partial{\cal L}_2},\ee
$${\cal R}^2=-16p^2{\cal L}_2{\cal L}_4^2+64p^2P({\cal L}_2,{\cal H}),$$
where 
$$ P({\cal L}_2,{\cal H})=(({\cal L}_2-\alpha-\beta)^2-4\alpha\beta)^q({\cal H}^2+4\omega^2{\cal L}_2)^p.$$
This shows that the symmetry algebra generated by ${\cal H}, {\cal L}_2, {\cal L}_4$ closes at finite order.

\medskip\noindent {\bf Case $p+q$ even}: Let 
\be \label{evenL34} {\cal L}_4=\frac{1}{i\sqrt{{\cal L}_2}}({\cal L}^- -{\cal L}^+),\ 
{\cal L}_3={\cal L}^+ +{\cal L}^-,\ee
where ${\cal L}^\pm$ are defined as in (\ref{oddL+-}).
Then we see from (\ref{plusbinom}), (\ref{minusbinom}) that ${\cal L}_3,{\cal L}_4$ are constants of the motion,
 polynomial in the momenta. Moreover, the identity (\ref{fundident1}) is still applicable, as are (\ref{raising}) and (\ref{lowering}).
Now we obtain
$$\{{\cal L}_2,{\cal L}_3\}=-4ip\sqrt{{\cal L}_2}{\cal L}^++4ip\sqrt{{\cal L}_2}{\cal L}^-=-4p{\cal L}_2{\cal L}_4,$$
$$\{{\cal L}_2,{\cal L}_4\}=\frac{1}{i\sqrt{{\cal L}_2}}[4ip\sqrt{{\cal L}_2}{\cal L}^+4ip\sqrt{{\cal L}_2}{\cal L}^-]=4p{\cal L}_3.$$
Similarly, since
\be\label{L4L41} {\cal L}_4^2=-\frac{1}{{\cal L}_2}\left[ ({\cal L}^+)^2-2{\cal L}^+{\cal L}^-+({\cal L}^-)^2\right],\ee
\be\label{L3L31}{\cal L}_3^2=\left[ ({\cal L}^+)^2+2{\cal L}^+{\cal L}^-+({\cal L}^-)^2\right].\ee
  we can re derive (\ref{L3^2}) again. Using (\ref{L+L-}) and (\ref{L4L41}) we re derive (\ref{L3L4bracket}). Hence the structure equations are exactly
(\ref{structure+}) again, even though the expressions for ${\cal L}_3,{\cal L}_4$ are different.

These structure results for classical TTW correspond to the highest order terms in the structure equations for quantum TTW, with the correspondence
$${\cal H}\sim H, \quad {\cal L}_2\sim L_2,\quad {\cal L}_3\sim -kL_3,\quad {\cal L}_4\sim L_4.$$
Furthermore, modulo multiplication by constants, the raising and lowering operators are associated with the classical  lowering and raising functions 
via $\Xi_\pm\sim {\cal L}^\mp$ and the quantum recurrences are associated with the classical factors via $K^-\sim Y$, $K^+\sim \overline{Y}$, $J^-\sim X$, $J^+\sim \overline{X}$. Multiplication or division of the quantum recurrence operators by $2n+a+b+1$ corresponds to multiplication or division of the classical raising and lowering symmetries by $\sqrt{{\cal L}_2}$.

\subsection{The classical constant of the motion ${\cal L}_5$}\label{L5}
As for the quantum TTW system with $k=1$, our method doesn't  give the generator of minimal order. In fact, for all rational $k$,  we can extend our 
symmetry algebra by adding a generator of order one less than ${\cal L}_4$. To see this, consider first the case
 $p+q$ odd. Then ${\cal L}_3=\frac{1}{i}({\cal L}^+-{\cal L}^-)$ is a polynomial in ${\cal L}_2$. Setting ${\cal L}_2=0$ in the expressions 
for $X, \overline{X}, Y,\overline{Y}$  we see that the constant term in ${\cal L}_3$ is  $2(-1)^{(p-q+1)/2} (\alpha-\beta)^q {\cal H}^p$. Thus
\be \label{L51} {\cal L}_5=\frac{{\cal L}_3-2(-1)^{(p-q+1)/2} (\alpha-\beta)^q {\cal H}^p}{{\cal L}_2}\ee
is a constant of the motion, polynomial in the momenta. From (\ref{L51}) we see that
\be \label{L52} {\cal L}_2{\cal L}_5={\cal L}_3-2(-1)^{(p-q+1)/2} (\alpha-\beta)^q {\cal H}^p,\ee
so taking the Poisson bracket with ${\cal L}_2$ on both sides of the equation, and using the relation $\{{\cal L}_2,{\cal L}_3\} =-4p{\cal L}_2{\cal L}_4$, 
we
find 
\be\label{L53} \{{\cal L}_2,{\cal L}_5\}= -4p{\cal L}_4.\ee
Then the identities we have already derived show that ${\cal L}_2,{\cal L}_5,{\cal H}$ generate a closed 
symmetry algebra that properly contains our original symmetry algebra.

If $p+q$ is even, then ${\cal L}_3={\cal L}^++{\cal L}^-$. Again ${\cal L}_3$ is  a polynomial in ${\cal L}_2$, but now with constant term 
$2(-1)^{(q-p)/2} (\alpha-\beta)^q {\cal H}^p$. Thus
\be\label{L54} {\cal L}_5=\frac{{\cal L}_3-2(-1)^{(q-p)/2} (\alpha-\beta)^q {\cal H}^p}{{\cal L}_2}\ee
is a constant of the motion, polynomial in the momenta.
For example, if $p=q=1$ then 
$${\cal L}_5=\frac{{\cal L}_3-2(\alpha-\beta){\cal H}}{{\cal L}_2},$$
is a 2nd order constant of the motion. Just as before ${\cal L}_2,{\cal L}_5, {\cal H}$ generate a closed symmetry algebra that properly contains our 
original algebra.

The construction we have given here is quite general and not at all restricted to the TTW system. All that is 
required is that the constant term in ${\cal L}_3$ be a polynomial in ${\cal H}$ alone. In all of the examples we know, this construction 
gives the generator of minimum order. However, we don't have a proof that the order is minmal for all $k$.

\section{The classical potential $V=k^2\delta^2 (x+iy)^{k-1}/(x-iy)^{k+1}$}
Consider the classical Hamiltonian
\be\label{Vclass} {\cal H} =p_x^2+p_y^2+k^2\delta^2 \frac{(x+iy)^{k-1}}{(x-iy)^{k+1}}\ee
where $k=p/q$.  This system is real in Minkowski space with coordinates $x_1=x,$ $y_1=-iy$. The corresponding quantum system has Hamiltonian
\be\label{Vquant}  H =\partial_x^2+\partial_y^2 +k^2\delta^2\frac{(x+iy)^{k-1}}{(x-iy)^{k+1}}\ee
The energy eigenvalue equation is $H\Psi=E\Psi$. In\cite{KMPog10} this system was shown to be both classically and quantum superintegrable, and in
\cite{KKM10c} the recurrence relation method was used to obtain the structure of the quantum symmetry algebra. 
Here, we revisit the classical problem and determine
the structure of the classical symmetry algebra. 

We use the fact that the Hamilton-Jacobi equation is additively separable in polar coordinates $x=r\cos\theta$, $y=r\sin\theta$ where $r=e^R$.
As in the previous example it is convenient to pass to variables $R$ and 
$\theta$. Then
$${\cal H}= {\cal L}_1=
\frac{(p^2_R+p^2_\theta +k^2\delta^2 e^{2ik\theta })}{ e^{2R}},\quad V=k^2\delta^2e^{2ik\theta-2R}$$
and there is is a constant of the motion 
$$ {\cal L}_2=p^2_\theta +k^2\delta^2 e^{2ik\theta},$$
responsible for the variable separation.

Using the usual prescription for obtaining the extra constant we  compute $M(R)=\frac12\int\frac{dR}{p_R}$,
 $N(\theta)=\frac12\int\frac{d\theta}{p_\theta}$. The new constant of the motion (not a polynomial) will be $M-N$.
If $k$ is an integer it is convenient to consider the solution such that  (in terms of the coordinates $R,\theta$)
we have 
$2ip\sqrt {{\cal L}_2}(M-N)=qA+pB$ as a constant of the motion
and 
$$
\sinh{ A}=\frac{i \sqrt{{\cal L}_2}}{ \delta k  }e^{-ik\theta},\ 
\cosh{ A}=\frac{ip_\theta}{ \delta k}e^{-ik\theta},$$
$$\sinh{ B}=\frac{i\sqrt{{\cal L}_2}}{ \sqrt{ {\cal H}}} e^{-R},\ 
\cosh{ B}=\frac{p_R}{ \sqrt{{\cal H}}}e^{-R}.$$
There is the identity $$e^{2R}{\cal H}=p_R^2+{\cal L}_2.$$
Using the same reasoning as for the classical TTW  case,
we have 
\be\label{expAV} e^A=\cosh A +\sinh A= {X}/{U},\quad e^{-A}=\cosh A -\sinh A= {\overline{X}}/{U},\ee
\be\label{expBV} e^B=\cosh B +\sinh B= {Y}/{S},\quad e^{-B}=\cosh B -\sinh B= {\overline{Y}}/{S},\ee
where
$$ X= (ip_\theta+i\sqrt{{\cal L}_2})e^{-ik\theta},\   
{\overline X}= (ip_\theta-i\sqrt{{\cal L}_2})e^{-ik\theta},$$
$$Y= (p_R+i\sqrt{{\cal L}_2})e^{-R},\ 
{\overline Y}= (p_R-i\sqrt{{\cal L}_2})e^{-R},$$
$$ U=k\delta,\quad S=\sqrt{{\cal H}}.$$
Then $e^{qA+pB}$ and $e^{-(qA+pB)}$ are constants of the motion, where
\be\label{expABV} e^{qA+pB}=(e^A)^q (e^B)^p=\frac{X^qY^p}{U^qS^p},\quad e^{-(qA+pB)}=(e^{-A})^q (e^{-B})^p=\frac{(\overline{X})^q
(\overline{Y})^p}{U^qS^p}.\ee The identity $e^{qA+pB}e^{-(qA+pB)}=1$ can be expressed as
\be\label{fundidentV} X^q(\overline{X})^qY^p(\overline{Y})^p=U^{2q}S^{2p}=(k\delta)^{2q}{\cal H}^p.\ee

  Using an analysis similar to  (\ref{plusbinom}), (\ref{minusbinom}), we see that, independent of $p+q$ being even or odd, we can set 
\be \label{oddL34V} {\cal L}_4=\frac{1}{\sqrt{{\cal L}_2}}({\cal L}^+ -{\cal L}^-),\ 
{\cal L}_3=({\cal L}^+ +{\cal L}^-),\ee
where
\be\label{oddL+-V} {\cal L}^+=X^qY^p,\quad {\cal L}^-=(\overline{X})^q(\overline{Y})^p.\ee
Then  ${\cal L}_3,{\cal L}_4$ are constants of the motion,
 polynomial in the  momenta. Moreover, the identity (\ref{fundidentV}) becomes
\be\label{fundident1V}{\cal L}^+{\cal L}^-=(k\delta)^{2q}{\cal H}^p.\ee

It is straightforward to verify the Poisson bracket computations 
\be\label{raisingV}\{{\cal L}_2,{\cal L}^+\}=q\frac{X^{q-1}Y^p}{U^qS^p}\{{\cal L}_2,X\}=2ip\sqrt{{\cal L}_2}{\cal L}^+,\ee
\be\label{loweringV}\{{\cal L}_2,{\cal L}^-\}=q\frac{(\overline{X})^{q-1}(\overline{Y})^p}{U^qS^p}\{{\cal L}_2,\overline{X}\}
=-2ip\sqrt{{\cal L}_2}{\cal L}^-.\ee
From these results we can obtain
$$\{{\cal L}_2,{\cal L}_4\}=\frac{1}{\sqrt{{\cal L}_2}}(2ip\sqrt{{\cal L}_2}{\cal L}^++ 2ip\sqrt{{\cal L}_2}{\cal L}^-)
=2ip{\cal L}_3,$$
$$\{{\cal L}_2,{\cal L}_3\}= 2ip\sqrt{{\cal L}_2}{\cal L}^+-2ip\sqrt{{\cal L}_2}{\cal L}^-=2ip{\cal L}_2{\cal L}_4.$$

Note that
\be\label{L4L4V} {\cal L}_4^2=\frac{1}{{\cal L}_2}\left[ ({\cal L}^+)^2-2{\cal L}^+{\cal L}^-+({\cal L}^-)^2\right],\
{\cal L}_3^2= ({\cal L}^+)^2+2{\cal L}^+{\cal L}^-+({\cal L}^-)^2.\ee
Thus,
\be\label{L3^2V} {\cal L}_3^2={\cal L}_2{\cal L}_4^2+4(k\delta)^{2q}{\cal H}^p.\ee

Note that ${\cal L}^-=(k\delta)^{2q}{\cal H}^p/{\cal L}^+$, so 
$$\{{\cal L}^+,{\cal L}^-\}=(k\delta)^{2q}\{{\cal L}^+,\frac{{\cal H}^p}{{\cal L}^+}\}=\frac{(k\delta)^{2q}}{{\cal L}^+}\{{\cal L}^+,{\cal H}^p\}=0$$
since ${\cal L}^+$ is a constant of the motion.

We find
$$\{{\cal L}_3,{\cal L}_4\}=\{{\cal L}^++{\cal L}^-,\frac{{\cal L}^+}{\sqrt{{\cal L}_2}}-\frac{{\cal L}^-}{\sqrt{{\cal L}_2}}\}=$$
$$-({\cal L}_2)^{-3/2}\{{\cal L}^+,{\cal L}_2\}{\cal L}^+ +({\cal L}_2)^{-3/2}\{{\cal L}^+,{\cal L}_2\}{\cal L}^- 
-({\cal L}_2)^{-3/2}\{{\cal L}^-,{\cal L}_2\}{\cal L}^++({\cal L}_2)^{-3/2}\{{\cal L}^-,{\cal L}_2\}{\cal L}^-$$
$$ +({\cal L}_2)^{-1/2}\{{\cal L}^+,{\cal L}^-\}+({\cal L}_2)^{-1/2}\{{\cal L}^-,{\cal L}^+\}$$
$$=\frac{2pi}{{\cal L}_2}[({\cal L}^+)^2-2{\cal L}^+{\cal L}^-+({\cal L}^-)^2].$$
Then, using (\ref{L4L4V}), we conclude that 
\be\label{L3L4bracketV}\{{\cal L}_3,{\cal L}_4\}=2pi{\cal L}_4^2.\ee
To list the structure equations in standard form we define ${\cal R}=-2pi{\cal L}_3$. Then the structure equations take the form
\be\label{structure+V} \{{\cal L}_2,{\cal L}_4\}={\cal R},\ \{{\cal L}_2,{\cal R}\}=-4p^2{\cal L}_2{\cal L}_4,\ \{{\cal L}_4,{\cal R}\}=4p^2{\cal L}_4^2,\ee
$${\cal R}^2=-4p^2{\cal L}_2{\cal L}_4^2-16p^2(k\delta)^{2q}{\cal H}^p.$$
This shows that the symmetry algebra generated by ${\cal H}, {\cal L}_2, {\cal L}_4$ closes at finite order.
As for the TTW case, these classical structure relations correlate with the quantum structure equations in \cite{KKM10c} (with the sign of the potential reversed) and give the classical version of recurrence relations.
\section{A system on the 2-sphere}
In \cite{KKM10c}  we considered a  quantum Hamiltonian  on the
two-sphere: 
\be \label{ham1} H=\partial ^2_\theta +\cot\theta \partial _\theta + \frac{1}{
\sin ^2\theta } 
\partial ^2_\varphi  + \frac{\alpha }{ \sin ^2\theta \cos^2k\varphi }\ee 
where $k=\frac{p}{ q}$ and $p,q$ are relatively prime positive integers, showed it was superintegrable, and determined the closure of its
 symmetry algebra via the recurrence relation approach.  Now we treat the classical analog
\be \label{ham2} {\cal  H}=p ^2_\theta  + \frac{1}{
\sin ^2\theta } 
p ^2_\varphi  + \frac{\alpha }{ \sin ^2\theta \cos^2k\varphi }.\ee 
In this case the variables are not in the standard form for our classical construction and we need to consider  the 
 separation variables $x_1,x_2$ as $ \varphi, \psi$ where  $x_1=\varphi$, $x_2=\psi$, $f_1=0$ , $f_2=1/\cosh^2\psi$,
$v_1=\alpha/\cos^2 k\varphi$. Then 
\be \label{ham3} {\cal  H}=\cosh^2\psi( p ^2_\psi  + 
p ^2_\varphi  + \frac{\alpha }{  \cos^2k\varphi })\ee 
and 
\be\label{L2S} {\cal L}_2=p ^2_\varphi  + \frac{\alpha }{  \cos^2k\varphi },\ee
so ${\cal H}=\cosh^2\psi(p_\theta^2+{\cal L}_2)$. The variables $\psi,\theta$ and their corresponding momenta are 
related by $\cosh\psi=1/\sin\theta$, $\sinh\psi=\cot\theta$ and $p_\psi=-\sin\theta\  p_\theta$. Then
$$M(\varphi,p_\varphi)=\frac12\int\frac{d\varphi}{p_\varphi}=\frac12\int\frac{d\varphi}{\sqrt{{\cal L}_2 -\frac{\alpha}{\cos^2 k\varphi}}},\
N(\psi ,p_\psi)=\frac12\int\frac{d\psi}{p_\psi}=\frac12\int\frac{d\psi}{\sqrt{\frac{{\cal H}}{\cosh^2\psi}-{\cal L}_2}},$$
so 
$$M=\frac{-iA}{2k\sqrt{{\cal L}_2}},\quad \sinh A= \frac{i\sqrt{{\cal L}_2}\sin  k\varphi}{\sqrt{{\cal L}_2-\alpha}},
\quad \cosh A=\frac{\cos k\varphi\ p_\varphi}{\sqrt{{\cal L}_2-\alpha}},$$
and
$$ N=\frac{iB}{2\sqrt{{\cal L}_2}},\quad \sinh B=\frac{-ik\sqrt{{\cal L}_2}\sinh\psi}{\sqrt{{\cal H}-{\cal L}_2}}=
\frac{-ik\sqrt{{\cal L}_2}\cot\theta}{\sqrt{{\cal H}-{\cal L}_2}},\quad
\cosh B=\frac{\cosh \psi\  p_\psi}{\sqrt{{\cal H}-{\cal L}_2}}=\frac{- p_\theta}{\sqrt{{\cal H}-{\cal L}_2}}.$$

From these expressions for $M$ and $N$ we see that
$$2ip\sqrt{{\cal  L}_2}[M-N]=qA+pB$$ is a constant of the motion. As usual,
\be\label{expAS} e^A=\cosh A +\sinh A= {X}/{U},\quad e^{-A}=\cosh A -\sinh A= {\overline{X}}/{U},\ee
\be\label{expBS} e^B=\cosh B +\sinh B= {Y}/{S},\quad e^{-B}=\cosh B -\sinh B= {\overline{Y}}/{S},\ee
where
$$ X= \cos k\varphi\ p_\varphi+i\sqrt{{\cal L}_2}\sin k\varphi,\   
{\overline X}= \cos k\varphi\ p_\varphi-i\sqrt{{\cal L}_2}\sin k\varphi,$$
$$Y=-p_\theta-ik\sqrt{{\cal L}_2}\cot \theta,\ 
{\overline Y}= -p_\theta+ik\sqrt{{\cal L}_2}\cot \theta,$$
$$ U=\sqrt{{\cal L}_2-\alpha},\quad S=\sqrt{{\cal H}-{\cal L}_2}.$$

Again, $e^{qA+pB}$ and $e^{-(qA+pB)}$ are constants of the motion, where
\be\label{expABS} e^{qA+pB}=(e^A)^q (e^B)^p=\frac{X^qY^p}{U^qS^p},\quad e^{-(qA+pB)}=(e^{-A})^q (e^{-B})^p=\frac{(\overline{X})^q
(\overline{Y})^p}{U^qS^p},\ee and the identity $e^{qA+pB}e^{-(qA+pB)}=1$ can be expressed as
\be\label{fundidentS} X^q(\overline{X})^qY^p(\overline{Y})^p=U^{2q}S^{2p}=P({\cal L}_2, {\cal H})=({\cal L}_2-\alpha)^q({\cal H}-{\cal L}_2)^p\ee
where $P$ is a polynomial in ${\cal L}_2$ and ${\cal H}$.

As in the example immediately preceding, we need consider only one case, independent of $p+q$ even or odd:
\be \label{L34S} {\cal L}_4=\frac{1}{\sqrt{{\cal L}_2}}({\cal L}^+ -{\cal L}^-),\ 
{\cal L}_3={\cal L}^+ +{\cal L}^-,\ee
where
\be\label{L+-S} {\cal L}^+=X^qY^p,\quad {\cal L}^-=(\overline{X})^q(\overline{Y})^p.\ee
Then  ${\cal L}_3,{\cal L}_4$ are constants of the motion,
 polynomial in the momenta. and 
\be\label{fundident1S}{\cal L}^+{\cal L}^-=P({\cal L}_2,{\cal H})=({\cal L}_2-\alpha)^q({\cal H}-{\cal L}_2)^p.\ee

Now,
\be\label{raisingS}\{{\cal L}_2,{\cal L}^+\}=q\frac{X^{q-1}Y^p}{U^qS^p}\{{\cal L}_2,X\}=-2ip\sqrt{{\cal L}_2}{\cal L}^+,\
\{{\cal L}_2,{\cal L}^-\}=q\frac{(\overline{X})^{q-1}(\overline{Y})^p}{U^qS^p}\{{\cal L}_2,\overline{X}\}
=2ip\sqrt{{\cal L}_2}{\cal L}^-,\ee
so
$$\{{\cal L}_2,{\cal L}_4\}=\frac{1}{\sqrt{{\cal L}_2}}(-2ip\sqrt{{\cal L}_2}{\cal L}^+- 2ip\sqrt{{\cal L}_2}{\cal L}^-)
=-2ip{\cal L}_3,$$
$$\{{\cal L}_2,{\cal L}_3\}= (-2ip\sqrt{{\cal L}_2}{\cal L}^++2ip\sqrt{{\cal L}_2}{\cal L}^-)
=-2ip{\cal L}_2{\cal L}_4.$$

Since
\be\label{L4L4S} {\cal L}_4^2=\frac{1}{{\cal L}_2}\left[ ({\cal L}^+)^2-2{\cal L}^+{\cal L}^-+({\cal L}^-)^2\right],\
{\cal L}_3^2= ({\cal L}^+)^2+2{\cal L}^+{\cal L}^-+({\cal L}^-)^2,\ee
we have 
$ {\cal L}_3^2-{\cal L}_2{\cal L}_4^2=4{\cal L}^+{\cal L}^-=4P({\cal L}_2,{\cal H})$, so
\be\label{L3^2S} {\cal L}_3^2={\cal L}_2{\cal L}_4^2+4P({\cal L}_2,{\cal H}).\ee

Further,  
$$\{{\cal L}^+,{\cal L}^-\} =\{{\cal L}^+, \frac{P}{{\cal L}^+}\}=  ({\cal L}^+)^{-1}\{{\cal L}^+,P\}=({\cal L}^+)^{-1}\left(\frac{\partial P}{\partial {\cal L}_2}\{{\cal L}^+,{\cal L}_2\}+\frac{\partial P}{\partial {\cal H}}\{{\cal L}^+,{\cal H}\}\right),$$
so
\be\label{L+L-S} \{{\cal L}^+,{\cal L}^-\}=2ip\sqrt{{\cal L}_2}\frac{\partial P}{\partial{\cal L}_2}.\ee
Lastly,
$$\{{\cal L}_3,{\cal L}_4\}=\{{\cal L}^++{\cal L}^-,\frac{{\cal L}^+}{\sqrt{{\cal L}_2}}-\frac{{\cal L}^-}{\sqrt{{\cal L}_2}}\}=$$
$$-({\cal L}_2)^{-3/2}\{{\cal L}^+,{\cal L}_2\}{\cal L}^+ +({\cal L}_2)^{-3/2}\{{\cal L}^+,{\cal L}_2\}{\cal L}^- 
-({\cal L}_2)^{-3/2}\{{\cal L}^-,{\cal L}_2\}{\cal L}^++({\cal L}_2)^{-3/2}\{{\cal L}^-,{\cal L}_2\}{\cal L}^-$$
$$ -({\cal L}_2)^{-1/2}\{{\cal L}^+,{\cal L}^-\}+({\cal L}_2)^{-1/2}\{{\cal L}^-,{\cal L}^+\}$$
\be\label{L3L4bracketS}=2ip{\cal L}_4^2-4ip\frac{\partial P}{\partial{\cal L}_2}.\ee
For standard form we set ${\cal R}=-2ip{\cal L}_3$. Then the structure equations take the form
\be\label{structure+S} \{{\cal L}_2,{\cal L}_4\}={\cal R},\ \{{\cal L}_2,{\cal R}\}=-4p^2{\cal L}_2{\cal L}_4,\ \{{\cal L}_4,{\cal R}\}=-4p^2{\cal L}_4^2+8p^2\frac{\partial P}{\partial{\cal L}_2},\ee
$${\cal R}^2=-4p^2{\cal L}_2{\cal L}_4^2-16p^2P({\cal L}_2,{\cal H}),$$
where 
$$ P({\cal L}_2,{\cal H})=({\cal L}_2-\alpha)^q({\cal H}-{\cal L}_2)^p.$$
Thus the symmetry algebra generated by ${\cal H}, {\cal L}_2, {\cal L}_4$ closes at finite order.

\section{The caged anisotropic oscillator revisited} The quantum caged anisotropic oscillator is defined by the Hamiltonian operator
\be\label{cagedoscillatorQ} H=\partial ^2_x+\partial ^2_y-\omega_1 ^2x^2-\omega_2
^2y^2+ 
\frac{\alpha}{x^2} + \frac{\beta}{ y^2},\ee
where $\omega _1=p\omega $ and $\omega _2=q\omega $ and $p$, $q$ are positive integers that
we assume are relatively prime. The classical version is
 \be\label{cagedoscillatorC} H=p^2_x+p ^2_y-\omega_1 ^2x^2-\omega_2
^2y^2+ 
\frac{\alpha}{x^2} + \frac{\beta}{ y^2},\ee
Marquette \cite{Marquette20101, Marquette20103}, using a ladder operator method,  was the first to work out the quantum structure 
relations for this system. We treated it again, from the recurrence relation point of
 view, in \cite{KKM10c} and will revisit it here to demonstrate the correspondence between the quantum recurrence relations and our 
classical construction. We do not claim novelty for the final results for this very simple but very important system, only the approach. 
This system separates in the Cartesian coordinates $x,y$ and either of the classical constants of the motion
$${\cal L}_1=p_x^2-\omega_1^2x^2+\frac{\alpha}{x^2},\quad
{\cal L}_2=p_y^2-\omega_2^2y^2+\frac{\beta}{y^2},$$
characterizes the separation, where  ${\cal H}={\cal L}_1+{\cal L}_2$.  

Constructing $M(x,p_x)$ which satisfies $M=\frac12\int dx/p_x$, we find a solution
$$ M=\frac{A}{4\omega_1},\ \sinh A= \frac{2\omega_1 x p_x}{ \sqrt{ {\cal L} ^2_1+4\omega ^2_1\alpha }}, \
\cosh A = \frac{2x^2\omega ^2_1+{\cal L}_1}{ \sqrt {L^2_1+4\omega ^2_1\alpha}}.$$
Similarly, $N(y,p_y)$ can be found as
$$ N=\frac{-B}{ 4\omega _2},\ \sinh B= \frac{-2\omega _2yp_y}{ \sqrt {L^2_2+4\omega ^2_2\beta}},\ 
\cosh B= \frac{2y^2\omega ^2_2+{\cal L}_2}{ \sqrt {L^2_2+4\omega ^2_2\beta}}.$$
Thus, $4pq\omega (M-N)=qA+pB$ is a constant of the motion, as are
\be\label{expABO} e^{qA+pB}=(e^A)^q (e^B)^p=\frac{X^qY^p}{U^qS^p},\quad e^{-(qA+pB)}=(e^{-A})^q (e^{-B})^p=\frac{(\overline{X})^q
(\overline{Y})^p}{U^qS^p},\ee and the identity $e^{qA+pB}e^{-(qA+pB)}=1$ can be expressed as
\be\label{fundidentO} X^q(\overline{X})^qY^p(\overline{Y})^p=U^{2q}S^{2p}=P({\cal L}_1,
 {\cal L}_2)=({\cal L} ^2_1+4\omega ^2_1\alpha )^q({\cal L} ^2_2+4\omega ^2_2\alpha )^p\ee
where $P$ is a polynomial in ${\cal L}_1$ and ${\cal L}_2$. In  particular,

$$ X= 2x^2\omega_1^2+{\cal L}_1+2\omega_1x p_x,\   
{\overline X}= 2x^2\omega_1^2+{\cal L}_1-2\omega_1x p_x,$$
$$Y=2y^2\omega_2^2+{\cal L}_2-2\omega_2y p_y,\   
{\overline Y}= 2y^2\omega_2^2+{\cal L}_2+2\omega_2y p_y,$$
$$ U=\sqrt {L^2_1+4\omega ^2_1\alpha},\quad S=\sqrt {L^2_2+4\omega ^2_2\beta} .$$

As usual, we set
\be\label{L+-O} {\cal L}^+=X^qY^p,\quad {\cal L}^-=(\overline{X})^q(\overline{Y})^p.\ee
In this degenerate case, ${\cal L}^+,{\cal L}^-$ are each constants of the motion, polynomial in the momenta, We set 
$${\cal L}_3= {\cal L}^++{\cal L}^-,\quad {\cal L}_4= {\cal L}^+-{\cal L}^-.$$
Now 
$$\{{\cal L}_1, {\cal L}^+\}=qX^{q-1}Y^p\{{\cal L}_1,X\}=-4q\omega_1{\cal L}^+,\ \{{\cal L}_1,{\cal L}^-\}=4q\omega_1{\cal L}^-,$$
$$\{{\cal L}_2,{\cal L}^+\}=pX^qY^{p-1}\{{\cal L}_2,Y\}=4p\omega_2{\cal L}^+,\{{\cal L}_2,{\cal L}^-\}=-4p\omega_2{\cal L}^-,$$
and, since ${\cal L}^+{\cal L}^-=P$, 
$$\{{\cal L}^+,{\cal L}^-\}=\{{\cal L}^+,\frac{P}{{\cal L}^+}\}=({\cal L}^+)^{-1}\{{\cal L}^+,P\}$$
$$=({\cal L}^+)^{-1}\left(\frac{\partial P}{\partial {\cal L}_1}\{{\cal L}^+,{\cal L}_1\}+\frac{\partial P}{\partial {\cal L}_2}\{{\cal L}^+,{\cal L}_2\}\right) =4q\omega_1\frac{\partial P}{\partial {\cal L}_1}-4p\omega_2\frac{\partial P}{\partial {\cal L}_2}.$$
Thus,
$$\{{\cal L}_1,{\cal L}_3\}=-4q\omega_1{\cal L}_4,\ \{{\cal L}_1,{\cal L}_4\}=-4q\omega_1{\cal L}_3,\ \{{\cal L}_2,{\cal L}_3\}=4p\omega_2{\cal L}_4,\ \{{\cal L}_2,{\cal L}_4\}=4p\omega_2{\cal L}_3,$$
$$\{{\cal L}_1,{\cal L}_2\}=0,\ \{{\cal L}_3,{\cal L}_4\}=8q\omega_1\frac{\partial P}{\partial {\cal L}_1}-8p\omega_2\frac{\partial P}{\partial {\cal L}_2}.$$
$$ {\cal L}_3^2-{\cal L}_4^2=4P,\quad P({\cal L}_1,{\cal L}_2)=({\cal L} ^2_1+4\omega ^2_1\alpha )^q({\cal L} ^2_2+4\omega ^2_2\alpha )^p.$$

\section{Extension of the generic potential on the 2-sphere}
A general Hamiltonian on the 2-sphere, separable in spherical coordinates can be taken in the standard form
$${\cal H}=\cosh^2\psi(p_\psi^2+p_\varphi^2+v_1(\varphi)+v_2(\psi)),\quad x=\varphi,\ y=\psi,\ f_1=0,\ f_2=\frac{1}{\cosh^2\psi}.$$
 Embedded in complex Euclidean 3-space with Cartesian coordinates, such 2-sphere systems can be written as
$$ H={\cal J}_1^2+{\cal J}_2^2+{\cal J}_3^2 +V({\bf s}),$$
where 
$${\cal J}_3=s_1p_{s_2}-s_2p_{s_1},\ {\cal J}_1=s_2p_{s_3}-s_3p_{s_2},\ {\cal J}_2=s_3p_{s_1}-s_1p_{s_3},\  s_1^2+s_2^2+s_3^3=1.$$
where $s_1=\sin\theta\cos\varphi$, $s_2=\sin\theta\sin\varphi$, $s_3=\cos\theta$ are the usual spherical coordinates. 
 The variables $\psi,\theta$ and their corresponding momenta are related by $\cosh\psi=1/\sin\theta$, $\sinh\psi=\cot\theta$ and $p_\psi=-\sin\theta\  p_\theta$.
Now we consider the superintegrable Hamiltonian system
\be \label{hamS9} {\cal H}=\cosh^2\psi [p^2_\psi +p^2_\varphi + \frac{\alpha }{\cos ^2k\varphi } + 
\frac{\beta }{\sin ^2k\varphi } + \frac{\gamma }{ \sinh^2\psi } ]\ee
$${\cal L}_2=p^2_\varphi + \frac{\alpha }{ \cos ^2k\varphi } + \frac{\beta }{ \sin ^2k\varphi }$$
Here $k=p/q$ where $p,q$ are relatively prime integers.
As shown in \cite{KKM10c}, 
 the functions that determine the extra constant of the motion are
$$M(\varphi ,p_\varphi )= \frac{i}{4k\sqrt{ {\cal L}_2}} {\cal A},\quad \sinh{\cal  A }= 
\frac{i({\cal L}_2\cos (2k\varphi )-\alpha +\beta )}{ \sqrt {({\cal L}_2-\alpha -\beta )^2-4\alpha \beta }},\ 
\cosh{\cal A}=\frac{\sqrt{{\cal L}_2}\sin (2k\varphi )p_\varphi }{ \sqrt {({\cal L}_2-\alpha -\beta )^2-4\alpha \beta }},$$
$$N(\psi ,p_\psi )= \frac{i}{ 4\sqrt {{\cal L}_2}}{\cal B},\quad \sinh{\cal B }= 
\frac{i({\cal L}_2\cosh(2\psi )+\gamma -{\cal H})}{ \sqrt {({\cal H}-{\cal L}_2-\gamma )^2-4{\cal L}_2\gamma }},\ 
\cosh{\cal B }= \frac{\sqrt{{\cal L}_2}\sinh(2\psi )p_\psi }
{ \sqrt{({\cal H}-{\cal L}_2-\gamma )^2-4{\cal L}_2\gamma }}.$$
(Here, we have corrected some misprints in the expressions listed there.) If $k=1$ this is the generic potential on the 2-sphere,
 a 2nd order superintegrable system.

Carrying out our usual construction, we have
\be\label{expAG} e^A=\cosh A +\sinh A= {X}/{U},\quad e^{-A}=\cosh A -\sinh A= {\overline{X}}/{U},\ee
\be\label{expBG} e^B=\cosh B +\sinh B= {Y}/{S},\quad e^{-B}=\cosh B -\sinh B= {\overline{Y}}/{S},\ee
where
$$ X= \sqrt{{\cal L}_2}\sin (2k\varphi) p_\varphi+i(\beta-\alpha+{\cal L}_2\cos(2k\varphi)),\   
{\overline X}= \sqrt{{\cal L}_2}\sin (2k\varphi) p_\varphi-i(\beta-\alpha+{\cal L}_2\cos(2k\varphi)),$$
$$Y=\sqrt{{\cal L}_2}\sinh (2\psi) p_\psi+i(\gamma-{\cal H}+{\cal L}_2\cosh(2\psi)),\    
{\overline Y}=\sqrt{{\cal L}_2}\sinh (2\psi) p_\psi-i(\gamma-{\cal H}+{\cal L}_2\cosh(2\psi)),$$
$$ U=\sqrt{({\cal L}_2-\alpha-\beta)^2-4\alpha\beta},\quad S=\sqrt{({\cal H}-{\cal L}_2-\gamma)^2-4\gamma{\cal L}_2}.$$
Then
\be\label{expABG} e^{qA+pB}=(e^A)^q (e^B)^p=\frac{X^qY^p}{U^qS^p},\quad e^{-(qA+pB)}=(e^{-A})^q (e^{-B})^p=\frac{(\overline{X})^q
(\overline{Y})^p}{U^qS^p},\ee  and  we have the identity 
\be\label{fundidentG} X^q(\overline{X})^qY^p(\overline{Y})^p=U^{2q}S^{2p}=P({\cal L}_2, {\cal H})=[({\cal L}_2-\alpha-\beta)^2-4\alpha\beta]^q
[({\cal H}-{\cal L}_2-\gamma)^2-4\gamma{\cal L}_2]^p\ee
where $P$ is a polynomial in ${\cal L}_2$ and ${\cal H}$.

\medskip\noindent {\bf Case $p+q$ odd}:  Let 
\be \label{oddL34G} {\cal L}_4=\frac{1}{\sqrt{{\cal L}_2}}({\cal L}^+ +{\cal L}^-),\ 
{\cal L}_3=\frac{1}{i}({\cal L}^+ -{\cal L}^-),\ee
where
\be\label{oddL+-G} {\cal L}^+=X^qY^p,\quad {\cal L}^-=(\overline{X})^q(\overline{Y})^p.\ee
Then  ${\cal L}_3,{\cal L}_4$ are constants of the motion,
 polynomial in the momenta and we have 
\be\label{fundident1G}{\cal L}^+{\cal L}^-=P({\cal L}_2,{\cal H}).\ee

We find 
\be\label{raisingG}\{{\cal L}_2,{\cal L}^+\}=q\frac{X^{q-1}Y^p}{U^qS^p}\{{\cal L}_2,X\}=4ip\sqrt{{\cal L}_2}{\cal L}^+,\ee
\be\label{loweringG}\{{\cal L}_2,{\cal L}^-\}=q\frac{(\overline{X})^{q-1}(\overline{Y})^p}{U^qS^p}\{{\cal L}_2,\overline{X}\}
=-4ip\sqrt{{\cal L}_2}{\cal L}^-,\ee
so
$$\{{\cal L}_2,{\cal L}_4\}=-4p{\cal L}_3,\quad
\{{\cal L}_2,{\cal L}_3\}=4p{\cal L}_2{\cal L}_4.$$
Using (\ref{L4L4}), (\ref{L3L3}) we obtain 
\be\label{L3^2G} {\cal L}_3^2=-{\cal L}_2{\cal L}_4^2+4P({\cal L}_2,{\cal H}).\ee
Further,  
$$\{{\cal L}^+,{\cal L}^-\} =\{{\cal L}^+, \frac{P}{{\cal L}^+}\}=  ({\cal L}^+)^{-1}\{{\cal L}^+,P\}=({\cal L}^+)^{-1}\left(\frac{\partial P}{\partial {\cal L}_2}\{{\cal L}^+,{\cal L}_2\}+\frac{\partial P}{\partial {\cal H}}\{{\cal L}^+,{\cal H}\}\right),$$
so
\be\label{L+L-G} \{{\cal L}^+,{\cal L}^-\}=-4ip\sqrt{{\cal L}_2}\frac{\partial P}{\partial{\cal L}_2},\ee
and
\be\label{L3L4bracketG}\{{\cal L}_3,{\cal L}_4\}=4p{\cal L}_4^2-8p\frac{\partial P}{\partial{\cal L}_2}.\ee
To list the structure equations in standard form we define ${\cal R}=-4p{\cal L}_3$. Then the structure equations take the form
\be\label{structure+G} \{{\cal L}_2,{\cal L}_4\}={\cal R},\ \{{\cal L}_2,{\cal R}\}=-16p^2{\cal L}_2{\cal L}_4,\ \{{\cal L}_4,{\cal R}\}=16p^2{\cal L}_4^2-32p^2\frac{\partial P}{\partial{\cal L}_2},\ee
$${\cal R}^2=-16p^2{\cal L}_2{\cal L}_4^2+64p^2P({\cal L}_2,{\cal H}),$$
where 
$$ P({\cal L}_2,{\cal H})=(({\cal L}_2-\alpha-\beta)^2-4\alpha\beta)^q(({\cal H}-{\cal L}_2-\gamma)^2-4\gamma{\cal L}_2)^p.$$
This shows that the symmetry algebra generated by ${\cal H}, {\cal L}_2, {\cal L}_4$ closes at finite order.

\medskip\noindent {\bf Case $p+q$ even}: Let 
\be \label{evenL34G} {\cal L}_4=\frac{1}{i\sqrt{{\cal L}_2}}({\cal L}^- -{\cal L}^+),\ 
{\cal L}_3={\cal L}^+ +{\cal L}^-,\ee
where ${\cal L}^\pm$ are defined as in (\ref{oddL+-G}). Then ${\cal L}_3,\ {\cal L}_4$ are polynomial constants of the motion. 
It can be verified that the structure equations are exactly the same as for $p+q$ odd.

Just as for the TTW system, in the case $k=1$ our method gives a 3rd order generator, whereas we know that, due to the
 multiseparability of the Hamilton-Jacobi equation, there is a 2nd order generator ${\cal L}_5$ independent of ${\cal L}_2, {\cal H}$.
 The method for finding this symmetry ${\cal L}_5$ for all rational $k$
and showing that it determines a larger symmetry algebra that closes is exactly analogous
 to that in Subsection \ref{L5}. The message is that the symmetries ${\cal L}^\pm$ determine everything.

\section{Discussion and conclusions}
We have found the classical analog of the  recurrence relation approach for verifying quantum superintegrability for 2D Hamiltonian systems that
 admit a separation of variables, and have demonstrated that this approach allows us give explicit expressions for the symmetries and to compute 
 structure equations for them. We have not proved that this method always yields the symmetry generators of lowest order, although that is the 
case for all of the examples we have checked. A major surprise for us was 
the simplicity of the structure equations. Indeed all of the structures in our various examples are essentially the same, except for the
 function $P({\cal L}_2,{\cal H})$ and the polynomial in ${\cal H}$ that occurs in the expression for ${\cal L}_5$.
It is clear that this method can be generalized to Hamiltonian systems in $>2$ dimensions that admit separation in subgroup type coordinates
 \cite{KKM10}. It is not clear what to do if the Hamiltonian system does not admit orthogonal variable separation.


\begin{thebibliography}{99}
\bibitem{TTW1} F.\ Tremblay, A.Turbiner and P.\ Winternitz. An infinite family of solvable and integrable quantum systems in the 
plane.  J.Phys.A: Math. Theor., {\bf 42},  (2009) 242001. 

\bibitem{TTW2} F.\ Tremblay, A.\ Turbiner and P.\ Winternitz. Periodic orbits for an infinite family of classical superintegrable 
systems.  J. Phys. A: Math. Theor. {\bf 43} (2010) 015202. 

\bibitem{KKM10} E. \ G.\ Kalnins, J.\ M.\ Kress and  W.\ Miller, Jr.     
Families of classical subgroup separable superintegrable systems.   J. Phys. A:
Math. Theor. 43 (2010) 092001.


\bibitem{KMPog10} E.\ G.\ Kalnins,  W.\ Miller Jr., and G.\ S.\ Pogosyan.
Superintegrability and higher order constants for classical and quantum systems,
Physics of Atomic Nuclei, (to appear) (2011)  
 arXiv:0912.2278v1 [math-ph]


\bibitem{KKM10a} E. \ G.\ Kalnins, J.\ M.\ Kress and  W.\ Miller, Jr.     
Superintegrability and higher order integrals for quantum systems, 
arXiv:1002.2665v1 [math-ph], J. Phys. A: Math. Theor. 43 (2010) 265205, 

\bibitem{KKM10b} E.\ G.\ Kalnins, J.\ M.\ Kress and W.\ Miller Jr.,.
Tools for verifying classical and quantum superintegrability. 
  arXiv:1006.0864v1 [math-ph],   SIGMA 6 (2010), 066.

\bibitem{KKM10c} E.\ G.\ Kalnins, J.\ M.\ Kress and W.\ Miller Jr.
A recurrence relation approach to higher order quantum superintegrability,   arXiv:1011.6548v3 [math-ph], (2011) submitted.  



\bibitem{CG}
C. Gonera. Note on Superintegrability of TTW model.
	arXiv:1010.2915v1 [math-ph], 2010.

\bibitem{MPT}
A. J. Maciejewski1, M. Przybylska, and A. V. Tsiganov.
On algebraic construction of certain integrable and
super-integrable systems, arXiv:1011.3249v1 [nlin.SI],  2010.


\bibitem{AAR} G.\ E.\ Andrews, R.\ Askey and R.\ Roy.  Special Functions.
Encyclopedia of Mathematics and
its Applications, Cambridge University Press, Cambridge, UK, 1999.


\bibitem{Evans2008a}
P.\ E.~Verrier and N.\ W.~Evans. A new superintegrable Hamiltonian.  J. Math. Phys.\ {\bf 49},  022902, 8 pages, arXiv:0712.3677, 2008.

\bibitem{KKMPog02}
E.\ G.\ Kalnins, J.\ M.\ Kress, W.\ Miller Jr., and G.\ S.\ Pogosyan. Complete sets of invariants for dynamical systems that admit separation of 
variables.  J. Math. Phys., {\bf 43}, 3592-3609,2002.

\bibitem{KKM20041}
E.\ G.\ Kalnins, J.\ M.\ Kress, W.\   Miller, Jr., 
Second  order superintegrable systems in conformally
flat spaces.  I: 2D classical structure theory,  J. Math. Phys. 46 (2005)
 053509.



\bibitem{Marquette20101} I.\ Marquette. Superintegrability and higher order polynomial algebras.  J.Phys A: Math. Theor. 43 135205 2010.

\bibitem{Marquette20103}I.\ Marquette. Construction of classical superintegrable systems with higher integrals 
of motion from ladder operators.
J. Math. Phys. 51 072903 2010.

\end{thebibliography}
\end{document}